\documentclass[pra,aps,groupedaddress,twocolumn,floatfix,nofootinbib]{revtex4}

\usepackage{graphicx} 
\usepackage[utf8]{inputenc}

\newcommand{\beq}{\begin{equation}}
\newcommand{\eeq}{\end{equation}}
\newcommand{\beqa}{\begin{eqnarray}}
\newcommand{\eeqa}{\end{eqnarray}}

\newcommand{\ket}[1]{\mbox{$ | #1 \rangle $}}
\newcommand{\bra}[1]{\mbox{$ \langle #1 | $}}

\def\opone{\leavevmode\hbox{\small1\normalsize\kern-.33em1}}

\begin{document}

\title{All entangled pure quantum states violate the bilocality inequality}
\author{Nicolas Gisin}
\affiliation{Groupe de Physique Appliqu\'ee, Universit\'e de Gen\`eve, CH-1211 Gen\`eve, Switzerland}
\author{Quanxin Mei}
\affiliation{Groupe de Physique Appliqu\'ee, Universit\'e de Gen\`eve, CH-1211 Gen\`eve, Switzerland}
\author{Armin Tavakoli}
\affiliation{Groupe de Physique Appliqu\'ee, Universit\'e de Gen\`eve, CH-1211 Gen\`eve, Switzerland}
\author{Marc Olivier Renou}
\affiliation{Groupe de Physique Appliqu\'ee, Universit\'e de Gen\`eve, CH-1211 Gen\`eve, Switzerland}
\author{Nicolas Brunner}
\affiliation{Groupe de Physique Appliqu\'ee, Universit\'e de Gen\`eve, CH-1211 Gen\`eve, Switzerland}

\date{\small \today}

\begin{abstract}
The nature of quantum correlations in networks featuring independent sources of entanglement remains poorly understood. Here, focusing on the simplest network of entanglement swapping, we start a systematic characterization of the set of quantum states leading to violation of the so-called ``bilocality'' inequality. First, we show that all possible pairs of entangled pure states can violate the inequality. Next, we derive a general criterion for violation for arbitrary pairs of mixed two-qubit states. Notably, this reveals a strong connection between the CHSH Bell inequality and the bilocality inequality, namely that any entangled state violating CHSH also violates the bilocality inequality. We conclude with a list of open questions.
\end{abstract} 

\maketitle

Quantum nonlocality, in the sense of violation of a Bell inequality, was considered as a mere curiosity---when not entirely ignored---during several decades after John Bell's seminal work \cite{Bell64}. Things changed dramatically in the early 1990's when Artur Ekert showed that nonlocality can be exploited to establish cryptographic keys between two remote observers \cite{Ekert91}. How could one ignore something useful for cryptography, especially in our information based society? Moreover, also in the early 1990's, experiments showed that the violation of Bell inequalities can be demonstrated over several kilometers using special optical fibers \cite{Rarity94} and even outside the controlled lab environment using standard telecom fibers \cite{TittelPRL98}. This led to rapid developments, both conceptually and for applications. Today, Bell inequality violation is routinely used in order to demonstrate the presence of entanglement in some physical system. This demonstrates quantumness beyond any doubt.

In the context of applications, quantum nonlocality led to the development of the field of device-independent quantum information processing \cite{BrunnerRMP14}, a way of processing information requiring no assumption about the details of the physical implementation, not even the dimension of the Hilbert space in which the quantum systems are represented. The measurement statistics suffice to guarantee security, for generating e.g. cryptographic keys \cite{Acin07}, or random numbers \cite{Pironio,Colbeck}. It is impressive that NIST has already made available online a beta version of a randomness beacon that will soon be offered to the public \cite{NISTrandomnessBeacon}. 

In the conceptual context, novel developments in quantum nonlocality have been inspired by experimental work on quantum networks. In such networks, there is not just one source of entanglement (the resource exploited for Bell inequality violation), but several sources distributing entanglement between different nodes, which can perform joint quantum measurements \cite{Kimble08}. This leads to strong correlations across the entire network. The understanding of such correlations is highly desirable, although still very limited at the moment. 

The simplest example of a joint quantum measurement is the so-called Bell Sate Measurement (BSM), a central ingredient in quantum teleportation \cite{QtelepPRL} and in entanglement swapping \cite{EntSwapZukowski93}. Formally, the BSM is represented by its four eigenvectors, namely the Bell states:
\beqa\label{BSM}
\ket{\phi^\pm}&=&\frac{1}{\sqrt{2}}(\ket{0,0}\pm\ket{1,1}) \\
\ket{\psi^\pm}&=&\frac{1}{\sqrt{2}}(\ket{0,1}\pm\ket{1,0})
\eeqa
hence referred to as a joint (or entangled) measurement. Since all Bell states are maximally entangled, their marginals 
are given by the maximally mixed state. Consequently, when one performs a BSM on independent qubits, all four results are equally likely, i.e. 25\% probability for each.

\begin{figure}[b!]
\includegraphics[width=\columnwidth]{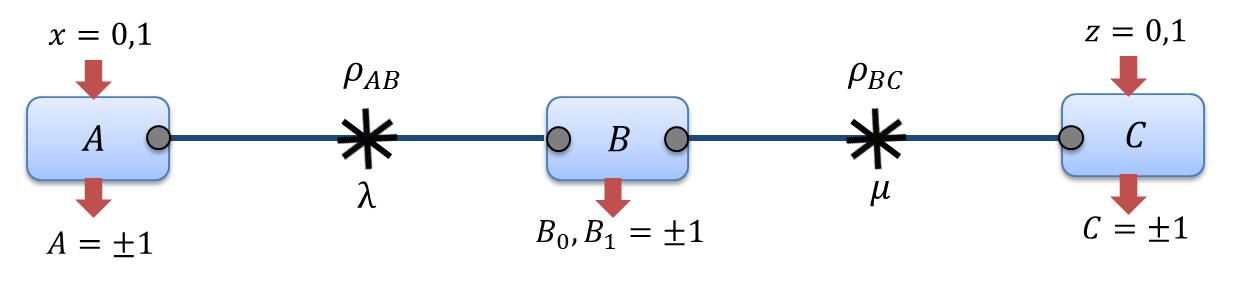}
\caption{Scenario of bilocality, the network we consider in this work. In the quantum setting, two independent sources distribute entangled states, $\rho_{AB}$ and $\rho_{BC}$, between three distant observers, Alice, Bob and Charlie. In order to compare the resulting quantum correlations to classical ones, we discuss $2$-local correlations obtained by two independent sources of shared classical random variables, $\lambda$ and $\mu$. For the bilocality inequality we consider, Alice and Charlie perform two dichotomic measurements, while Bob performs a fixed measurements with four possible outcomes. In the quantum setting, Bob's measurement is taken to be the Bell State Measurement.\label{fig_principle}}
\end{figure}

Figure 1 illustrates the simplest quantum network, with only three observers and two sources. This is the scenario we consider in this letter. In the standard analysis of this scenario, i.e. following Bell locality, one would contrast the correlations achievable with quantum resources, e.g. two sources of entangled pairs and the BSM in the middle, with classical resources, i.e. all three parties share some common local hidden variable (LHV). Note that ``local hidden variable'' is the old terminology, going back to EPR \cite{EPR35} and Bell \cite{Bell64}. Nowadays one refers to shared randomness, a terminology closer to cryptography, though technically synonymous. Hence, all three parties---named Alice, Bob and Charlie---would share a common classical random variable. 

However, looking at Fig. 1, it is arguably much more natural to contrast quantum correlations with classical correlations achievable via two independent sources of shared randomness. More precisely, Alice and Bob would share some variable $\lambda$ (originating from the source between them), while Bob and Charlie would share another variable $\mu$ (originating from the second source). Importantly the variables $\lambda$ and $\mu$ should be uncorrelated, as the two sources are independent. This independence assumption is very natural, given that the quantum network of Fig. 1 features two fully independent sources of entanglement.
There is thus no reason to assume that $\lambda$ and $\mu$ are correlated. And this very natural assumption changes everything!

This new scenario has been studied under the name of 2-locality (2- because of the two sources) or merely bilocality. More formally 2-local correlations are characterized as follows. Consider that Alice receives measurement setting (or input) $x$, while Bob gets input $y$, and Charlie $z$. Upon receiving their inputs, each party should provide a measurement result (an output), denoted $A$ for Alice, $B$ for Bob, and $C$ for Charlie. In this context, the observed statistics is said to be 2-local when
\beqa 
p(ABC|xyz) =\!\!\! \int  d \lambda   d \mu \, q_1(\lambda) q_2(\mu) 
p(A|x \lambda) \, p(B|y \lambda \mu) \, p(C|z \mu)   \nonumber
\eeqa
where $\lambda $ and $\mu$ are the independent shared random variables distributed according to the densities $q_1(\lambda)$ and $q_2(\mu)$, respectively. The set of 2-local correlations (i.e. the set of all correlations of the above form) is non-convex \cite{bilocality10}, rendering its analysis challenging. In particular, in order to efficiently characterize the 2-local set, non-linear Bell inequalities are required. Note that this in stark contrast to the set of Bell-local (or 1-local) correlations which is convex and can thus be fully characterized by linear Bell inequalities \cite{BrunnerRMP14}. 

In Refs \cite{bilocality10,bilocality12}, first non-linear inequalities that allow one to efficiently capture 2-local correlations (better than any linear inequality) were derived. Here we focus on an inequality presented in \cite{bilocality12}, which we will refer to as the bilocality inequality (for simplicity). Consider that Alice and Charlie receive binary inputs, $x=0,1$ and $z=0,1$, and must give binary outputs, denoted $A_x=\pm1$ and $C_z=\pm1$, respectively. The middle party Bob always performs the same measurement (hence receives no input $y$) with 4 possible outcomes, as e.g. the BSM. Denote Bob's outcome by two bits $B_0 = \pm 1$ and $B_1 = \pm 1$. 
The bilocality inequality reads:
\beq\label{bilocIneq}
S_{biloc}\equiv \sqrt{|I|}+\sqrt{|J|}\le2
\eeq
where 
\beqa
I&\equiv&  \langle (A_0+A_1)B_0(C_0+C_1) \rangle \label{I}\\
J&\equiv&  \langle  (A_0-A_1)B_1(C_0-C_1) \rangle  \label{J}.
\eeqa
The bracket $\langle \cdot \rangle$ denotes the expectation value of many experimental runs.

Interestingly this inequality can be violated by certain quantum correlations \cite{bilocality12}, that would have to be considered local in the usual Bell approach (i.e. when all three parties could have common shared randomness). In particular, consider the case where Alice-Bob, as well as Bob-Charlie, share a noisy Bell state (with visibility $V$), a so-called Werner state, of the form $\rho = V \ket{\phi^+} \bra{\phi^+} + (1-V) \frac{\openone}{4}$. Conditioned on one outcome of Bob's BSM, the state shared by Alice and Charlie is again a Werner state, but with lower visibility $V^2$. The bilocality inequality can be violated whenever $V^2>1/2$. This is in strong contrast with the usual Bell approach, where in order to detect quantum nonlocality, one would require a visibility $V > 1/\sqrt{2}$ using the CHSH \cite{CHSH} Bell inequality \footnote{Note that one could do marginally better ($V \simeq 0.705$) by using an inequality introduced by V\'ertesi \cite{vertesi}.}, while for visibilities up to $V \simeq 0.682$ the Werner state admits a LHV model \cite{Hirsch16} and can thus not violate any Bell inequality\footnote{Notice that this does not allow one to reveal the nonlocality of a Werner state $\rho$ with $V\leq 1/\sqrt{2}$ by distributing two copies of $\rho$ in the considered network and violate the bilocality inequality (see discution in \textit{conclusions}). However, it does constitute a significant advantage as compared to entanglement swapping experiments based on the CHSH Bell inequality. }.

The above results demonstrated the relevance of the 2-locality approach for detecting quantum correlations in networks. This triggered further research, with the derivation of novel non-linear inequalities and the exploration of more sophisticated networks; see e.g. \cite{fritz,branciard,tavakoli, Henson,Mukherjee14,Nloc,chaves,tavakoli2,tavakoli3}. However, the extent of quantum correlations in networks remains poorly understood. This is precisely the goal of the present work, where we start a systematic characterization of the class of quantum states leading to violation of the bilocality inequality (\ref{bilocIneq}).


%

\emph{All pairs of pure entangled states.}--- 
We start our analysis by considering that both sources emit pure entangled states. Denote $\ket{\psi_{AB}}=c_0\ket{00}+c_1\ket{11}$ and $\ket{\phi_{BC}}=q_0\ket{00}+q_1\ket{11}$ the two normalized (two-qubits) pure states shared by Alice and Bob and by Bob and Charlie, respectively, written in the Schmidt basis, with real and positive coefficients $c_j$ and $q_j$. Note that if these Schmidt bases would differ from the computational basis in which the BSM (\ref{BSM}) is written, then it would suffice to add local unitary rotations on each qubit to recover the case we discuss here.
Define $c=2c_0c_1$ and $q=2q_0q_1$; $\ket{\psi_{AB}}$ ($\ket{\phi_{BC}}$) are entangled whenever $c>0$ ($q >0$). Note that we can restrict to two-qubit entangled states here. If the states are of larger dimension, Alice, Bob and Charlie can first project them onto qubit sub-spaces, hence our setting is fully general for the case of two pure states \cite{Gisin91}.

Let Alice's inputs correspond to projective measurements in the Z-X plane of the Bloch sphere. Thus each can be characterized by an angle $\pm\alpha$ (with respect to the Z axis). The observable corresponding to the first input reads $\vec{a} \cdot \vec{\sigma}$, where $\vec{a} = (\sin(\alpha), 0 , \cos(\alpha))$ and $\vec{\sigma} = (\sigma_x, \sigma_y, \sigma_z)$ denotes the vector of Pauli matrices. Similarly for Charlie we have angles $\pm\gamma$. 
Bob performs the usual BSM. 
For all $x,z=0,1$ one gets:
\beqa
\langle A_xB_0C_z \rangle &=& \langle \big(\cos(\alpha)\sigma_z+(-1)^x \sin(\alpha)\sigma_x\big)\otimes(\sigma_z\otimes\sigma_z)\nonumber\\
& & \otimes \big(\cos(\gamma)\sigma_z+(-1)^z \sin(\gamma)\sigma_x\big) \rangle_{\psi_{AB}\otimes\phi_{BC}}\nonumber\\
&=&\cos(\alpha)\cos(\gamma) \,.
\eeqa
Hence $I=4\cos(\alpha)\cos(\gamma)$. A similar calculation gives $J=4\sin(\alpha)\sin(\gamma) c q$.

Maximizing expression (\ref{bilocIneq}) with respect to $\alpha$ and $\gamma$ leads to:
\beqa \label{sett_opt}
\cos(\alpha)&=&\cos(\gamma)=\frac{1}{\sqrt{1+cq}}
\eeqa
and the maximum takes the value:
\beqa
S_{biloc}^{max}&=&\sqrt{4\cos(\alpha)\cos(\gamma)}+\sqrt{4\sin(\alpha)\sin(\gamma) cq}\nonumber\\ \label{Scq}
&=&2\sqrt{1+cq} \,.
\eeqa
Accordingly, for all possible pairs of entangled pure states, i.e. when $c>0$ and $q>0$, we get violation the standard bilocality inequality (\ref{bilocIneq}) and thus non-bilocal correlations.

Note that if $\ket{\psi_{AB}}=\ket{\phi_{BC}}$, then the optimal settings $\alpha$ and $\gamma$ for bilocality are the same as the optimal settings for the CHSH inequality. Furthermore $S_{biloc}^{max}$ takes the same value as the maximum CHSH value for $\ket{\psi_{AB}}$ \cite{Gisin91}. 

Interestingly, if the states differ, then Alice's optimal settings depend on the state $\ket{\phi_{BC}}$ shared by Bob and Charlie, and similarly Charlie's optimal settings depend on $\ket{\psi_{AB}}$, as can be seen from Eq. (\ref{sett_opt}). 

Note that if one would now consider noisy states of the form $V_{AB} \ket{\psi_{AB}} \bra{\psi_{AB}} + (1-V_{AB}) \openone / 4$ and similarly for $V_{BC} \ket{\phi_{BC}} \bra{\phi_{BC}} + (1-V_{BC}) \openone / 4$, then one can characterize the critical visibilities ($V^{biloc}_{AB}$ and $V^{biloc}_{BC}$), i.e. the minimum visibilities for which violation of the bilocality inequality is still possible, which are in general related. More precisely, one finds that the product of the critical visibilities (for bilocality) is smaller than the product of the visibilities for Bell locality (i.e. 1-locality): $V^{biloc}_{AB}  V^{biloc}_{BC}=\frac{1}{1+cq}\le V^{loc}_{AB} V^{loc}_{BC}=\sqrt{\frac{1}{1+c^2}}\sqrt{\frac{1}{1+q^2}}$, with equality holding only when $c=q$, i.e. when the two states are equal, $\ket{\psi_{AB}}=\ket{\phi_{BC}}$.

\emph{Criterion for arbitrary pairs of mixed states.}---
We now move to mixed states, and start our analysis with the case of two-qubit density matrices. Let 
\beqa  \nonumber
\rho_{AB}=\frac{1}{4}(\opone+\vec{m}_A \cdot \vec{\sigma}\otimes\opone+
\opone\otimes\vec{m}_B \cdot \vec{\sigma} +\sum_{ij}t_{ij}^{AB} \sigma_i\otimes\sigma_j)
\eeqa
be the state shared by Alice and Bob, expressed in the Pauli basis; here the vector $\vec m_A$ ($\vec m_B$) represents the Bloch vector of Alice's (Bob's) reduced state, while $t_{ij}^{AB}$ (with $i,j \in \{x,y,z\}$) is the correlation matrix. Similarly we express $\rho_{BC}$, the state shared by Bob and Charlie, in the Pauli basis. 

Alice's settings are represented by Bloch vectors $\vec a$ and $\vec a'$, and similarly for Charlie $\vec c$ and $\vec c^{\hspace{0.5mm}'}$. Assume Bob performs a BSM in a well chosen basis to be defined below. 
The quantity $I$ in Eq. (\ref{I}) can be expressed as follows:
\beqa
I&=&Tr[(\vec a+\vec a') \cdot \vec\sigma\otimes\sigma_z\otimes\sigma_z\otimes(\vec c+\vec c^{\hspace{0.5mm}'}) \cdot \vec\sigma \, \rho_{AB}\otimes\rho_{BC}]\nonumber\\
&=&Tr[(\vec a+\vec a') \cdot \vec\sigma\otimes\sigma_z \, \rho_{AB}] \, Tr[\sigma_z\otimes(\vec c+\vec c^{\hspace{0.5mm}'})\cdot \vec\sigma \, \rho_{BC}]\nonumber\\
&=&\sum_i(a_i+a'_i)t_{iz}^{AB} \, \sum_k t_{3k}^{BC} (c_k+c_k')\label{zx} \,.
\eeqa
Using the polar decomposition, the correlation matrix can be written as $t^{AB}=U^{AB}R^{AB}$, where $U^{AB}$ is a unitary matrix and $R^{AB}=\sqrt{t^{AB \dagger} \, t^{AB} } \geq 0$. Denote $\xi_1\ge\xi_2\ge \xi_3 \ge 0$ the three non-negative eigenvalues of $R^{AB}$. Similarly denote $\zeta_1\ge\zeta_2\ge\zeta_3\ge0$ the non-negative eigenvalues of the corresponding matrix $R^{BC}$. 

This allows us to characterize Bob's BSM. Specifically, the Bell states (as given in Eqs (\ref{BSM})) has been defined such that the Z and X Bloch directions (on the first subsystem, connected to Alice) are given by the eigenvectors of the matrix $R^{AB}$ corresponding to the two largest eigenvalues, $\xi_1$ and $\xi_2$, respectively. Similarly we use $R^{BC}$ for aligning the second subsystem of Bob, connected to Charlie.  Note that the Z and X axes Bob uses with Alice may differ from those he uses with Charlie, i.e. Bob may have to apply different unitaries to the two qubits he shares with Alice and with Charlie before performing a standard BSM.

Next our goal is to maximize $S_{biloc}$ with respect to the Bloch vectors $\vec a,\vec a',\vec c$ and $\vec c^{\hspace{0.5mm}'}$. It is clear that they should lie within the two-dimensional subspace spanned by the two eigenvectors with largest eigenvalues: $\vec a=(\sin\alpha, 0, \cos\alpha)$, $\vec a'=(\sin\alpha', 0,\cos\alpha')$, $\vec c=(\sin\gamma,0, \cos\gamma)$ and $\vec c'=(\sin\gamma', 0,\cos\gamma')$. The maximum is easily found by imposing $\partial_\alpha S=0$, $\partial_{\alpha'} S=0$, $\partial_\gamma S=0$ and $\partial_{\gamma'} S=0$. One finds $\alpha'=-\alpha$, $\gamma'=-\gamma$ and
\beqa
\cos\alpha&=&\cos\gamma=\sqrt{\frac{\xi_1\zeta_1}{\xi_1\zeta_1+\xi_2\zeta_2}} \,,
\eeqa
and the maximal violation of the bilocality inequality 
\beq\label{Smax}
S_{biloc}^{max}=2\sqrt{\xi_1\zeta_1+\xi_2\zeta_2} \,.
\eeq
Consequently, a pair of states $\rho_{AB}$ and $\rho_{BC}$ can violate the bilocality inequality (\ref{bilocIneq}) if and only if 
$\xi_1\zeta_1+\xi_2\zeta_2>1$. Note that for the case of two pure states considered previously, $\xi_1=\zeta_1=1$, $\xi_2=2c_0c_1=c$ and $\zeta_2=2q_0q_1=q$; hence (\ref{Smax}) reduces to (\ref{Scq}), as it should.

The above criterion is analogous to the Horodecki criterion for violation of the CHSH Bell inequality \cite{Horodecki95}. In fact, there is a direct connection between the two criteria. According to the Horodecki criterion the maximal CHSH value for $\rho_{AB}$ is given by $ S_{AB}^{max} = 2 \sqrt{\xi_1^2 + \xi_2^2} = 2 ||\vec{\xi}||$ where $\vec{\xi} = (\xi_1,\xi_2)$. Similarly, for $\rho_{BC}$ we have $ S_{BC}^{max} = 2 \sqrt{\zeta_1^2 + \zeta_2^2} = 2 || \vec{\zeta}||$ with $\vec{\zeta} = (\zeta_1,\zeta_2)$. From Eq. (\ref{Smax}) it follows that
\beqa \label{eq12}
S_{biloc}^{max} = 2 \sqrt{ \vec{\xi} \cdot \vec{\zeta} } 
 \leq  2\sqrt{||\vec{\xi} || \, || \vec{\zeta} ||} =   \sqrt{S_{AB}^{max} \, S_{BC}^{max} }
\eeqa
Hence, violation of the bilocality implies that either $\rho_{AB}$ or $\rho_{BC}$ (or both) must violate CHSH. Moreover, when the two states are the same, i.e. $\rho_{AB}=\rho_{BC}= \rho$, the criterion of Eq. (\ref{Smax}) reduces to the Horodecki criterion. This is easily seen from Eq. (\ref{eq12}), where the inequality becomes an equality when the vectors $\vec{\xi}$ and $\vec{\zeta}$ are the same. Therefore, CHSH violation implies violation of the bilocality inequality in the sense that 
\beqa \label{equiv}
\rho \textrm{ violates CHSH}  \,\, \rightarrow \,\,  \rho \otimes \rho \textrm{ violates } S_{biloc}  \,.
\eeqa
Note that under the assumption that Bob performs the BSM, the reverse link also holds. In this case activation of nonlocality is thus impossible for two-qubit entangled states using the bilocality inequality (see discussion below). Note also that the connection (\ref{equiv}) holds true for arbitrary bipartite mixed states $\rho$, not only for two-qubit states \cite{armin_new}.

\emph{Conclusion.}---
In quantum networks involving several independent sources of entangled states, it is natural to contrast the obtained quantum correlations with ``classical'' correlations that can be realized using independent sources of shared randomness between the observers. Indeed, this picture is arguably a natural generalization to networks of John Bell's original intuition \cite{Bell64,BellSpeakable}. In the simplest case, i.e. with two independent sources as in entanglement swapping, we analyzed the standard bilocality inequality and proved that all pairs of entangled pure states can violate it, in analogy to the case of the CHSH-Bell inequality which can be violated by any pure entangled state. Moving to mixed entangled states, we then derived a general criterion for violation of the bilocality inequality, providing a natural extension of the Horodecki criterion for violation of CHSH. In particular this reveals a strong connection between CHSH and the bilocality inequality, namely that any entangled state violating CHSH can also be used to demonstrate violation of the bilocality inequality. 


While the results presented in this letter were obtained analytically, we conclude with a list of open questions that we could so far tackle only numerically:
\begin{enumerate}
\item Here we assumed that Bob performs a BSM, defined in local basis depending on the shared entangled states. One may expect that this is always optimal, which is confirmed numerically for any pair of pure states. However, numerical evidence suggests that there are cases, in which one or both states are mixed, for which no BSM is optimal. So far, we could not find any structure in the optimal joint measurements and leave it for future work. 
\item The bilocality inequality (\ref{bilocIneq}) used here assumes a scenario in which Bob has no choice of input and 4 possible outcomes. However, an inequality formally identical to (\ref{bilocIneq}) is also valid for the scenario in which Bob has a choice between two inputs with binary outcomes: it suffices to label $B_0$ and $B_1$ the outcomes corresponding to the two inputs, respectively. The bilocal bound of the inequality remains the same (because classically Bob could always compute and output both the value of $B_0$ and of $B_1$). However, quantum mechanically, Bob's two joint measurements may be incompatible, leading possibly to larger violations. We could confirm this possibility, though only numerically so far.

\item It would be interesting to generalize the present results to the case of more sophisticated networks, such as star networks \cite{tavakoli} with an arbitrary number of branches. 
\item A central open question is the possibility to activate the nonlocality of certain entangled quantum states---admitting a LHV model in the usual Bell scenario---by placing several copies of them in a network. While such effect is possible even when considering the standard Bell approach \cite{Cavalcanti11} (see also Ref.\cite{Klobus}), intuition suggests that the notion of $N$-locality should be very useful in this context. However, no examples have been reported so far. Here, we have proven that activation is impossible for the bilocality inequality when Bob performs the BSM. We also performed intensive numerical search considering more general measurements for Bob. The results suggest that activation is impossible for the bilocality inequality. A formal proof of this statement would be desirable. A counter-example would be even more interesting. 
\end{enumerate}

\small

\emph{Acknowledgments.}--- 
This work was supported by the Swiss national science foundation (SNSF 200021-149109 and Starting grant DIAQ), and the European Research Council (ERC-AG MEC).

\end{document}